# Spin-density-wave transition in monolayer-trilayer $La_3Ni_2O_7$ single crystals

Mingxin Zhang[1,2#], Jie Dou[3,4#], Di Peng[5,6#], Cuiying Pei[1], Qi Wang[1,7], Yi Zhao[1], Chao Xiong[3], Shuo Li[3,4], Jun Luo[3,4], Juefei Wu[1], Lingxiao Zhao[1], Qing Zhang[1,8], Jie Yang[3,4], Yulin Chen[1,7,9], Jinkui Zhao[10,2,3], Wenge Yang[5,6], Hanjie Guo[2*], Qiaoshi Zeng[5,6*] Rui Zhou[3,4*], Yanpeng Qi[1,7,8*]

1. State Key Laboratory of Quantum Functional Materials, School of Physical Science and Technology, ShanghaiTech University, Shanghai 201210, China
2. Songshan Lake Materials Laboratory, Dongguan 523808, Guangdong, China
3. Institute of Physics, Chinese Academy of Sciences, and Beijing National Laboratory for Condensed Matter Physics, Beijing 100190, China
4. China School of Physical Sciences, University of Chinese Academy of Sciences, 100190, Beijing, China
5. Center for High Pressure Science and Technology Advanced Research, Shanghai 201203, China
6. Shanghai Key Laboratory of Material Frontiers Research in Extreme Environments (MFree), Institute for Shanghai Advanced Research in Physical Sciences (SHARPS), Shanghai 201203, China
7. ShanghaiTech Laboratory for Topological Physics, ShanghaiTech University, Shanghai 201210, China
8. Shanghai Key Laboratory of High-resolution Electron Microscopy, ShanghaiTech University, Shanghai 201210, China
9. Department of Physics, Clarendon Laboratory, University of Oxford, Parks Road, Oxford OX1 3PU, UK
10. School of Physical Sciences, Great Bay University, Dongguan 523808, China

# These authors contribute equally to this work.

* Correspondence should be addressed to Y.P.Q. (qiyp@shanghaitech.edu.cn) or R.Z.( rzhou@iphy.ac.cn) or Q.S.Z. (zengqs@hpstar.ac.cn) or H.J.G (hjguo@sslab.org.cn)

## ABSTRACT

**The recent discovery of high-temperature superconductivity in pressurized Ruddlesden-Popper nickelates stimulated intense research into their correlated electron physics. Establishing the diversity of ground states across different Ruddlesden-Popper phases is crucial for elucidating the superconducting mechanisms in these nickelates. Motivated by the recent report of superconductivity in hybrid 1212-type $La_5Ni_3O_{11}$, we synthesized and investigated the long-range-ordered hybrid 1313-type $La_3Ni_2O_7$. In contrast to its bilayer counterpart, the 1313-type $La_3Ni_2O_7$ exhibits characteristic semiconducting behavior at ambient pressure, displaying a distinct anomaly at 170 K. This behavior is consistently evidenced by measurements of both magnetic susceptibility and specific heat. $^{139}La$ nuclear magnetic resonance spectroscopy unambiguously indicates a spin-density-wave transition occurring at 170 K. High-pressure electrical transport measurements demonstrate the induction of metallization under pressure, yet reveal no discernible traces of superconductivity up to 65 GPa. Our findings establish hybrid 1313-type $La_3Ni_2O_7$ as a new member of the Ruddlesden-Popper nickelate family exhibiting a distinct spin-density-wave transition, and offers a new platform for investigating the interplay among crystal structure, electronic orders, and superconductivity in hybrid nickelates.**

# I.INTRODUCTION

The discovery of high-temperature superconductivity ($T_c \approx 80$ K) in pressurized $La_3Ni_2O_7$ single crystals has revitalized research on nickelate superconductors[1-34]. Subsequent observation of superconductivity in other Ruddlesden-Popper (RP) phases — including bilayer $La_2PrNi_2O_7$, trilayer $R_4Ni_3O_{10}$($R$ = La, Pr) and monolayer-bilayer $La_5Ni_3O_{11}$[35-45] — have established diverse platforms for theoretical and experimental study. Significantly, thin film engineering has enabled stabilization of superconductivity at ambient pressure through strain manipulation[46-53], enabling investigations of the electronic structure and magnetic properties in the superconducting state [54, 55].

Despite considerable progress, the superconducting phase in $La_3Ni_2O_7$ remains contentious. This controversy stems primarily because hybrid 1313-type $La_3Ni_2O_7$, featuring alternating monolayer-trilayer stacking, also exhibits superconductivity near 80 K[56] . Unraveling the origin of high-$T_c$ superconductivity in $La_3Ni_2O_7$ faces significant challenges due to poorly controlled single-crystal growth. The principal obstacles are: (i) Incomplete phase separation between the bilayer (2222) and monolayer-trilayer (1313) polymorphs, resulting from overlapping synthesis window [56-58]. These persistent phase impurities critically impede subsequent intrinsic property measurements. (ii) Restricted synthesis pressure: single-crystal growth is confined to a narrow oxygen pressure range (3-5 bar), unlike others RP nickelates (e.g., $La_4Ni_3O_{10}$, $LaNiO_3$) [56, 57, 59-63]. This inherent limitation fundamentally degrades crystal quality by promoting structural imperfections and chemical inhomogeneity, thereby obscuring intrinsic electronic behavior.

In this work, we achieve the growth of high-quality hybrid 1313-type $La_3Ni_2O_7$ single crystals by utilizing elevated oxygen pressures to suppress the formation of the competing bilayer 2222 phase during crystallization. Comprehensive characterization by single-crystal and powder X-ray diffraction, and atomic-resolution scanning transmission electron microscopy (STEM) confirms the phase purity and long-range crystallographic order of the as-grown single crystals. Nuclear magnetic resonance (NMR) measurements definitively identify a spin-density-wave (SDW) transition at 170 K in hybrid 1313-type $La_3Ni_2O_7$ single crystals. Furthermore, our investigation reveals that the application of pressure effectively suppresses the weakly semiconducting ground state, inducing a pressure-driven metallization. By establishing a reliable single-crystal platform and systematically characterizing its electronic response to hydrostatic pressure, this work provides critical insight for resolving the persistent controversy surrounding the high-$T_c$ superconducting phase in $La_3Ni_2O_7$.

These advances lay the foundation for elucidating the microscopic mechanism of superconductivity in this correlated nickelate system.

## II. RESULTS AND DISCUSSION

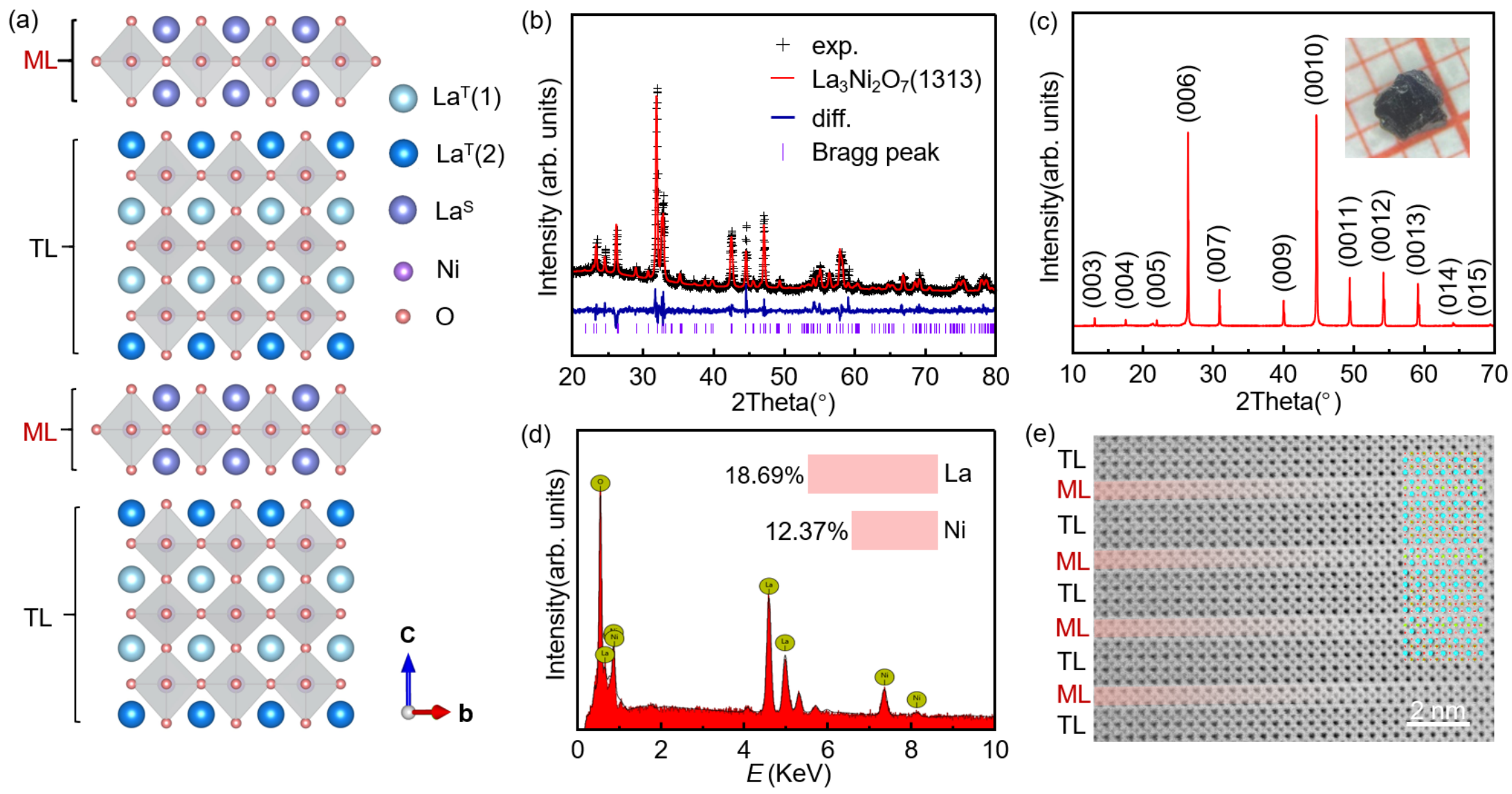

**FIG. 1. Crystal structure and characterization of hybrid 1313-type $La_3Ni_2O_7$ single crystals**. (**a**) The crystal structure of hybrid 1313-type $La_3Ni_2O_7$ single crystal. $La^S$ and $La^T$ denote La atoms in the monolayer and trilayer blocks, respectively. The bond angle of Ni–O–Ni along the *a*, *b*, and *c*-axis directions is 180°. (b) Powder X-ray diffraction pattern of hybrid 1313-type $La_3Ni_2O_7$ single crystal. (**c**) Single crystal X-ray diffraction pattern of hybrid 1313-type $La_3Ni_2O_7$ single crystal along the *ab* plane. The inset shows a photo of hybrid 1313-type $La_3Ni_2O_7$ single crystal. (**d**) Chemical composition of hybrid 1313-type $La_3Ni_2O_7$ single crystal taken by the energy-dispersive X-ray spectroscopy. (**e**) Annular bright-field scanning transmission electron microscopy images of hybrid 1313-type $La_3Ni_2O_7$ single crystal.

The high-quality single crystals of hybrid RP phase $La_3Ni_2O_7$ were grown using the optical floating zone method (See Supplementary Methods for detail). Structural characterization via powder X-ray diffraction (PXRD) confirms the as-grow crystal adopts the 1313-type with the *Cmmm* space group, featuring an alternating sequence of monolayer and trilayer perovskite blocks stacked along the *c*-axis (Fig. 1(a)). In contrast to the tilted $NiO_6$ octahedra in bilayer $La_3Ni_2O_7$ (2222) or trilayer $La_4Ni_3O_{10}$ (3333) phases [1, 36], the Ni-O-Ni angle along the *c*-axis in hybrid 1313-type $La_3Ni_2O_7$ is symmetry-constrained to 180°[56, 57, 59].

The tetragonal Ni-O plane of hybrid 1313-type $La_3Ni_2O_7$ under ambient pressure is analogous to the $La_4Ni_3O_{10}$ in the superconductivity state under pressure[36, 37], and the infinity $Nd_{1-x}Sr_xNiO_2$ thin film at ambient pressure[64]. Rietveld refinement of the PXRD yields the following lattice parameters for hybrid 1313-type $La_3Ni_2O_7$ : $a$ = 5.4376 Å, $b$ = 5.4786 Å, $c$ = 20.3328 Å, and α = β = γ = 90°(Fig. 1(b)). The high crystalline quality and orientation are further verified by single crystal XRD in the *ab* plane, which reveals exclusively (00*l*) Bragg peaks (Fig. 1(c)). Scanning electron microscope (SEM) images display a characteristic step-terrace surface, indicative of its layered crystal structure (Fig. S1). Finally, energy-dispersive X-ray spectroscopy (EDS) analysis confirms chemical composition (Fig. 1(d)), quantifying the La : Ni atomic ratio as approximately 1.51, which is consistent with the stoichiometric 3:2 ratio expected for hybrid 1313-type $La_3Ni_2O_7$.

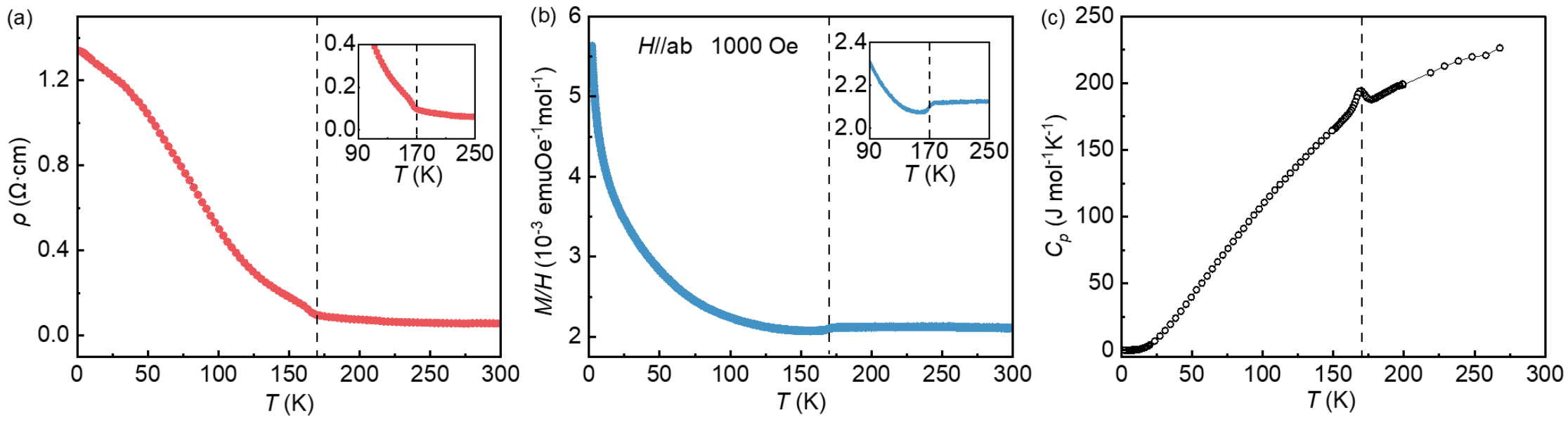

**FIG. 2. Transport properties of hybrid 1313-type $La_3Ni_2O_7$ single crystals.** (**a**) Temperature dependent resistivity of hybrid 1313-type $La_3Ni_2O_7$ single crystal. (**b**) Temperature-dependent magnetic susceptibility of hybrid 1313-type $La_3Ni_2O_7$ single crystal with an applied field of 0.1 T. (**c**) Temperature-dependent heat capacity of hybrid 1313-type $La_3Ni_2O_7$ single crystal.

The crystalline quality and atomic-scale order of as-grown hybrid 1313-type $La_3Ni_2O_7$ single crystal were further investigated using annular bright-field scanning transmission electron microscopy (ABF-STEM) and $^{139}$La-NMR spectroscopy. Atomic-resolution STEM imaging along the [110] projection reveals a perfectly ordered alternating sequences of monolayer-trilayer perovskite blocks, extending coherently over tens of nanometers in hybrid 1313-type $La_3Ni_2O_7$ single crystals without detectable stacking faults or phase intergrowths (Fig. 1(e)). This long-range structural order is corroborated by EDS element mapping, which clearly distinguishes the spatial distributions of La and Ni atoms (Fig. S2), with nickel and lanthanum positions sharply segregated in accordance with the expected crystal structure.

Within hybrid 1313-type $La_3Ni_2O_7$ structure featuring alternating single and triple $NiO_2$ layers (Fig. 1(a)), the La atoms occupy three distinct sites: $La^S$ resides in the monolayer, $La^T(1)$ lies within the trilayer, and $La^T(2)$ is located at the trilayer periphery, with an atomic occupancy ratio of 1: 1: 1. Given $^{139}$La nucleus with $I$ = 7/2, each site generates seven NMR transitions, so one expects three central transitions accompanied by three sets of satellite peaks. The $^{139}$La-NMR spectrum acquired at 206 K (Fig. S3), exhibits three groups of satellite peaks. This is distinct from the previous reports on $La_3Ni_2O_7$ (2222)[65] where only one set of satellite peaks was observed. Analysis of satellite spacings yields the apparent quadrupole frequencies of 1.31 MHz, 2.30 MHz, and 4.46 MHz. Owing to the single crystal's irregular morphology, the applied magnetic field was not perfectly perpendicular to the $c$-axis. The NQR measurements indicate that actual $\nu_Q$ = 5.15 MHz for the $La^T(2)$ site exhibiting maximal quadrupole frequency (Fig. S4). From the $\nu_Q$ of the $La^T(2)$ site, the $\nu_Q$ of other La sites can be derived (see Supplementary Information and Table S1). Comparison with previous reports (Table S1) allows assignment of the three sets of satellite peaks to $La^S$, $La^T(1)$, and $La^T(2)$ sites, as labeled in Fig. S5. Since the central peak is in the middle of the two satellite peaks, we can assign each component of the central line accordingly (see Fig. S5). The fitting of the central line using Lorentz functions yields the integrated intensity ratios $La^S$ : $La^T(1)$ : $La^T(2)$ = 1 : 1.2 : 0.82, congruent with stoichiometric proportions. In addition to the three peaks corresponding to these three La sites, we observed a minor peak adjacent to the central line of $La^S$ (see Fig. S5). One possible explanation is that the small peak originates from the residual bilayer phase within our sample. Nevertheless, no peak with $\nu_Q \approx$ 5.9 MHz, which corresponds to the quadrupole frequency of $La_3Ni_2O_7$ (2222)[66-68], was observed in both the NMR and NQR spectra(see Fig. S3 and S4), indicating that the bilayer phase is negligible in our sample. Then, it is possible that this peak could be associated with the La site in close proximity to an oxygen vacancy[15]. Collectively, these results confirm the fully ordering sequences of the alternating monolayers and trilayers in hybrid 1313-type $La_3Ni_2O_7$ single crystals, setting the stage for precise investigations into its physical properties under both ambient pressure and high pressure.

The electrical resistivity, magnetic susceptibility, and heat capacity of hybrid 1313-type $La_3Ni_2O_7$ single crystals were systematically measured, as summarized in Fig. 2. Unlike the insulating monolayer $La_2NiO_4$ and metallic trilayer $La_4Ni_3O_{10}$[60, 69], hybrid 1313-type $La_3Ni_2O_7$ exhibits semiconducting behavior manifested by monotonic increasing in resistivity upon cooling (Fig. 2(a)). A pronounced anomaly emerges at ~ 170 K, signaling the onset of a density-wave (DW) transition. Concomitant with the opening of the DW gap, the resistivity undergoes a rapid enhancement below this critical temperature. Fig. 2(b) displays the temperature-dependent magnetic susceptibility measured with an applied field of 0.1 T. Upon cooling from room temperature, the magnetic susceptibility exhibits a gradually decreases until the DW transition region. A marked suppression is observed near 170 K, followed by an upturn at lower temperatures. This susceptibility profile bears a notable resemblance to that of trilayer $La_4Ni_3O_{10}$[36, 60, 61, 70], suggesting a possible common origin rooted in the trilayer structural motif. However, the systematic alternation of monolayer ($La_2NiO_4$) and trilayer ($La_4Ni_3O_{10}$) blocks in hybrid 1313-type $La_3Ni_2O_7$ appears to enhance paramagnetic character compared to pure trilayer $La_4Ni_3O_{10}$ systems, indicating a significant modification of the interlayer magnetic coupling. The bulk nature of the DW transition is unequivocally corroborated by heat capacity measurements. As shown in Fig. 2(c), heat capacity exhibits a pronounced peak at a similar temperature (~170 K), providing thermodynamic evidence for the phase transition.

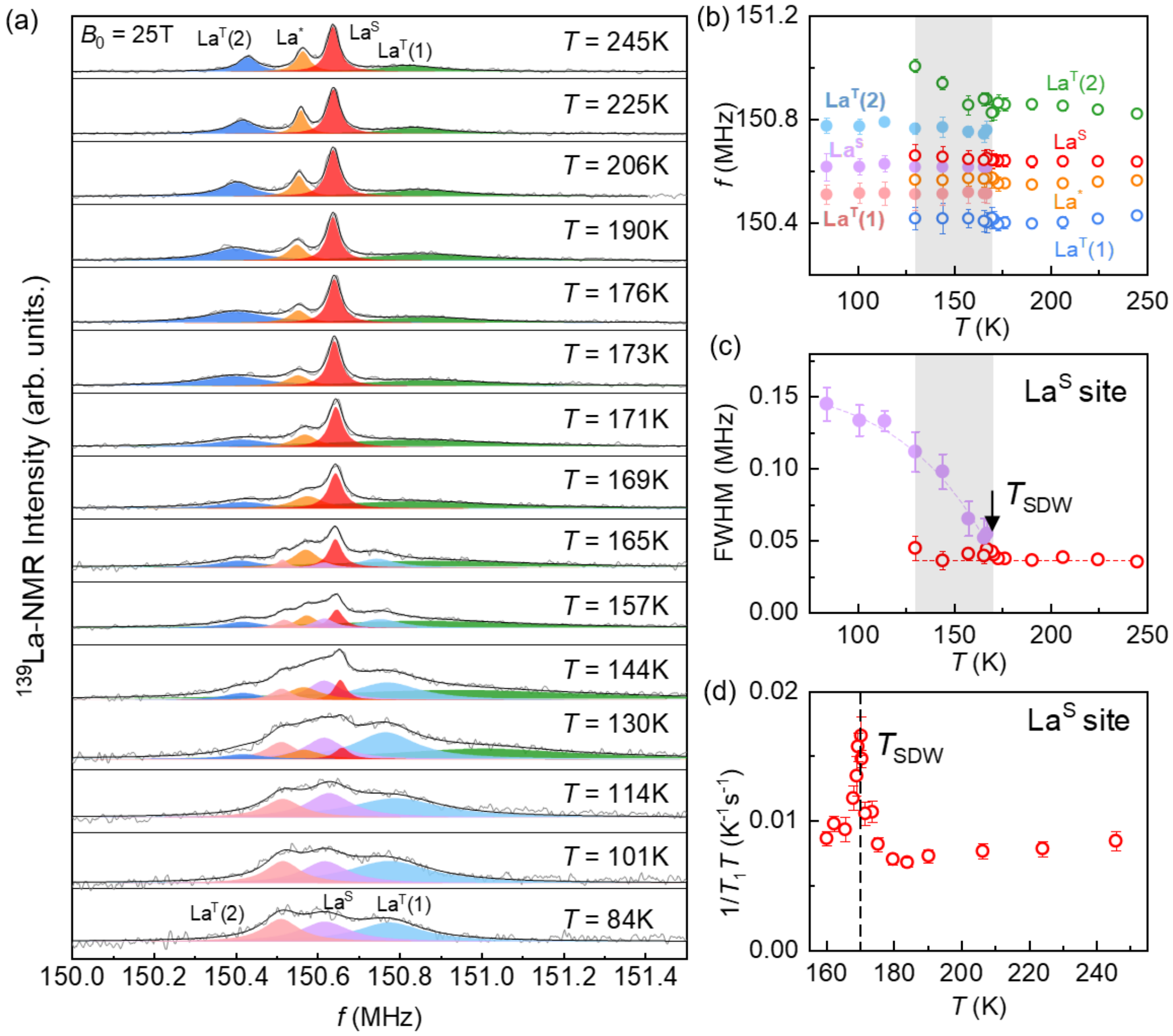

**FIG. 3. Spin-density-wave transition of hybrid 1313-type $La_3Ni_2O_7$ single crystals.** (**a**) The temperature dependence of $^{139}$La-NMR spectra of the central lines. The line designated as La$^*$ might be associated with the La site in close proximity to an oxygen vacancy[71]. The solid lines are fitted using Lorentz functions. (**b**) The temperature dependence of frequencies of each line in the central line of $^{139}$La-NMR spectra obtained. (**c**) The temperature dependence of FWHM of the central line corresponding to the La$^S$ site. The open red circles denote the FWHM of the red peak in (**a**), which corresponds to the paramagnetic phase. The solid purple circles denote the FWHM of the purple peak in (**a**), which corresponds to the magnetic phase. The dashed lines are the guide to the eye. (**d**) Temperature dependence of $1/T_1$ measured at the La$^S$ site. The black arrow and dashed line represent $T_{SDW}$. The gray area represents the coexistence region of the paramagnetic and magnetic phase. The error bars in peak position and FWHM represent the standard deviation in fitting the NMR spectra. The error bar in $T_1$ is the standard deviation in fitting the nuclear magnetization recovery curve.

To determine the fundamental order parameter of the DW transition, whether it is primarily of charge-density-wave (CDW) or SDW character, we performed a combined investigation using temperature-dependent electron diffraction patterns, single crystal X-ray diffraction, and $^{139}$La-NMR investigations. Representative electron diffraction patterns acquired at 300 K and 95 K are shown in Figs. S6(a)-(b). They exhibit identical Bragg reflections, showing no evidence of superlattice spots associated with a CDW-driven periodic lattice distortion. Given the local-probe nature of electron diffraction, potential modulated structures were further examined via single crystal X-ray diffraction on micron-scale single crystals. Reconstructed reciprocal-space maps (*h0l* planes) at 300 K and 100 K, presented in Fig. S6(c)-(d), reveal no detectable satellite reflections. This result definitively precludes the presence of dominant, long-range periodic lattice modulations, thereby ruling out a conventional CDW as the primary origin of the DW transition.

In contrast, the NMR measurements provide direct evidence for a spin-related order. Fig. 3(a) displays the temperature evolution of the central lines of $^{139}$La-NMR spectra, which manifests progressive change of the NMR spectra below $T \sim 170$ K, signaling SDW order. Below $T_{SDW} \sim 170$ K, not only does the broadening occur, but a narrow small peak at 150.64 MHz persists from high temperatures down to $T = 144$ K and then vanishes below $T = 130$ K, suggesting a coexistence of paramagnetic and magnetically ordered phases between 130 K and 170 K. Within this regime, the spectra are well fitted by using seven-peak Lorentz functions, contrasting with three-peak Lorentz functions are used at lower temperatures (see Fig. 3(a)). The fittings of NMR spectra were in good agreement with the experiment results. The temperature dependence of peak positions and full widths at half maximum (FWHM) can be deduced as shown in Figs. 3(b)-3(c). In the coexistence region, an abrupt change in the NMR frequencies is observed. Meanwhile, the FWHM is nearly temperature independent in the paramagnetic phase, whereas it shows an increase upon cooling below $T_{SDW} \sim 170$ K. All these findings suggest that the observed SDW transition is a first-order transition. Moreover, we observed that the spin–lattice relaxation rate divided by temperature, $1/T_1T$, exhibits a peak near 170 K (see Fig. 3(d)). The similar temperature dependence of FWMH and $1/T_1T$ to that of the $La_3Ni_2O_7$ (2222) [65, 67, 68] unambiguously demonstrates an SDW formation in hybrid 1313-type $La_3Ni_2O_7$ at $T_{SDW} \approx 170$ K.

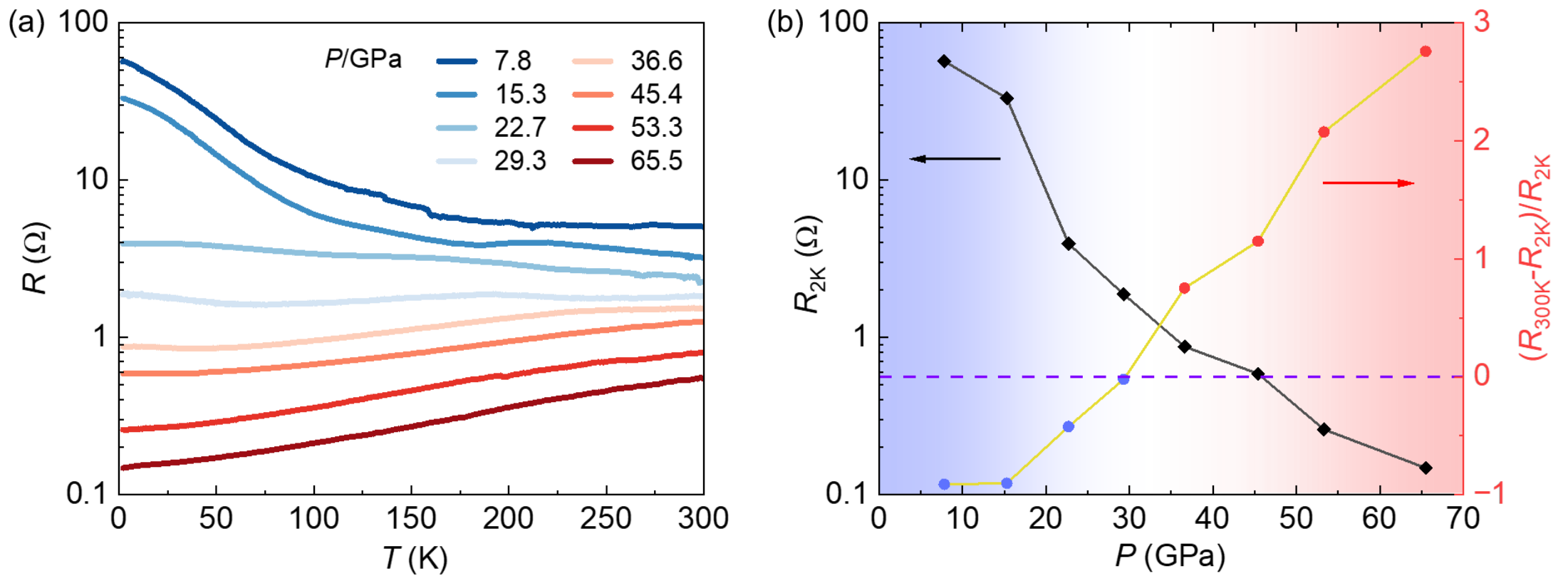

**FIG. 4. The pressure effect of hybrid 1313-type $La_3Ni_2O_7$ single crystals.** (**a**) Temperature dependent resistance of $La_3Ni_2O_7$ single crystals at 7.8-65.5 GPa using helium as pressure transmitting medium. (**b**) Pressure-dependence of the resistance at 2 K and the normalized resistance change, $(R_{300K}\text{-}R_{2K})/R_{2K}$, illustrating the semiconductor-to-metal transition. Blue and red symbols denote the semiconducting and metallic regions, respectively.

The high-quality hybrid 1313-type $La_3Ni_2O_7$ single crystal exhibits a well-defined SDW transition. Motivated by the emergency of high-$T_c$ superconductivity in related RP phases — including both conventional (e.g., bilayer $La_3Ni_2O_7$, $La_4Ni_3O_{10}$) and hybrid structures (e.g., $La_5Ni_3O_{11}$)—we probe the electronic response of hybrid 1313-type $La_3Ni_2O_7$ under pressure via electrical transport measurements. Electrical transport measurements were performed using both solid (cubic boron nitride) and gas (helium) as pressure transmitting medium. Temperature-dependent resistance under pressure spanning 0.5-50.1 GPa is presented in Fig. S7. At the lowest pressure (0.5 GPa), the resistance increases monotonically upon cooling, consistent with semiconducting behavior at ambient pressure. Progressive application of pressure systematically suppresses this semiconducting tendency. Crucially, no superconducting transition is observed up to 50.8 GPa in our samples. We note that an upturn in resistance emerges at low temperatures. This feature, reminiscent of the behavior observed in hybrid 1313 thin films synthesized at ambient pressure [72], may be attributed to residual disorder effects such as local oxygen vacancies or internal strain. The previous work on nickelate superconductors has demonstrate the sensitivity of superconductivity to hydrostatic pressure[1, 2, 45]. Fig. 4(a)-(b) presents the detailed evolution of resistance under hydrostatic pressure using helium as the pressure transmitting medium. As pressure increases, the weakly semiconducting behavior is progressively suppressed, and the low-temperature upturn gradually diminishes. Despite this pressure-induced metallization, superconductivity fails to emerge throughout the investigated pressure range.

This absence of superconductivity aligns with recent theorical works that the electronic structure of hybrid 1313-type $La_3Ni_2O_7$ is incompatible with high-$T_c$ superconductivity[73]. Corroborating evidence emerges from strained hybrid 1313-type $La_3Ni_2O_7$ thin films grown under ambient-pressure, which also exhibits metallic behavior without any sign of superconductivity down to the lowest temperature[72]. Collectively, these findings demonstrate the superconductivity near 80 K originates exclusively in bilayer (2222) $La_3Ni_2O_7$ phase. From a structural perspective, the $T_c$ in 1212-type $La_5Ni_3O_{11}$ (64 K) is already reduced compared to bilayer $La_3Ni_2O_7$ (80 K). The inherently unfavorable coordination geometry in hybrid 1313-type $La_3Ni_2O_7$ is expected to impose further suppression relative to the trilayer $La_4Ni_3O_{10}$, potentially suppressing superconductivity entirely.

## III. CONCLUSION

We have successfully synthesized high-quality, phase-pure hybrid 1313-type $La_3Ni_2O_7$ single crystals. Building on this structural foundation, we performed comprehensive transport measurements under both ambient and hydrostatic pressure conditions. Correlated analysis of transport and NMR data reveals the formation of a SDW ground state below $T_{SDW} \approx 170$ K. High-pressure resistance measurements up to 65 GPa demonstrate systematic suppression of the semiconducting behavior yet show no evidence of superconductivity. This absence of superconductivity is consistent with recent experimental findings in strained thin films and with theoretical predictions for hybrid 1313-type electronic structure. This work provides key insights into the superconducting phase formation in $La_3Ni_2O_7$ and establishes a robust materials platform for probing the fundamental interplay among crystal structure, electronic orders, and superconductivity in correlated nickelates.

## ACKNOWLEDGEMENTS

This work was supported by the National Key R&D Program of China (Grant No. 2023YFA1607400) and the National Natural Science Foundation of China (Grant Nos. 52272265, 12474018). The work at Institute of Physics, Chinese Academy of Sciences was supported by the National Key Research and Development Projects of China (Grant Nos. 2023YFA1406103 and 2024YFA1611302), the National Natural Science Foundation of China (Grant Nos. 12374142, 12304170, and U23A6003), and the CAS Superconducting Research Project (Grant No. SCZX-01011). Qiaoshi Zeng and Di Peng acknowledge the support of the Shanghai Key Laboratory of Material Frontiers Research in Extreme Environments, China (grant no. 22dz2260800), and the Shanghai Science and Technology Committee, China (grant no. 22JC1410300). Hanjie Guo acknowledges the support from the Guangdong Basic and Applied Basic Research Foundation (Grant No. 2022B1515120020). A portion of this work was carried out at the Synergetic Extreme Condition User Facility (SECUF, https://cstr.cn/31123.02.SECUF). The authors thank the support from Analytical Instrumentation Center (# SPSTAIC10112914), SPST, ShanghaiTech University.